\documentclass[twocolumn]{aastex631}
\usepackage{CJK}
\newcommand{\uatnum}[1]
{\href{http://vocabs.ands.org.au/repository/api/lda/aas/the-unified-astronomy-thesaurus/current/resource.html?uri=http://astrothesaurus.org/uat/#1}{#1}}
\usepackage{multirow}
\usepackage{booktabs}
\usepackage{glossaries}
\usepackage{soul}
\usepackage[nodayofweek]{datetime}
\newdateformat{monthyearday}{%
  \THEYEAR\ \monthname[\THEMONTH] \twodigit{\THEDAY}}
\received{\monthyearday 2024 February 29}
\revised{\monthyearday 2024 August 9}
\accepted{\monthyearday 2024 August 12}
\published{}
\submitjournal{The Astronomical Journal}

\graphicspath{{./}{}}
\shorttitle{Eps Eri Inner Ring Limit}
\shortauthors{P.M. et al.}

\newacronym{AU}{AU}{astronomical Unit [1.5e11 m]}  
\newacronym{pc}{pc}{parsec}
\newacronym{mas}{mas}{milliarcsecond}
\newacronym{nm}{nm}{nanometer}
\newacronym{CTE}{CTE}{coefficient of thermal expansion}
\newacronym{sqarc}{$as^2$}{square arcsecond}
\newacronym{MAD}{MAD}{Median absolute deviation}

\newacronym{smc}{SMC}{Small Magellanic Cloud}
\newacronym{lmc}{LMC}{Large Magellanic Cloud}
\newacronym{ism}{ISM}{interstellar medium}
\newacronym{mw}{MW}{Milky Way}
\newacronym{epseri}{$\epsilon$ Eri}{Epsilon Eridani}
\newacronym{deleri}{$\delta$ Eri}{Delta Eridani}
\newacronym{EKB}{EKB}{Edgeworth-Kuiper Belt}

\newacronym{CFR}{CFR}{Complete Frequency Redistribution}

\newacronym{nasa}{NASA}{National Aeronautics and Space Agency}
\newacronym{esa}{ESA}{European Space Agency}
\newacronym{omi}{OMI}{\textit{Optical Mechanics Inc.}}
\newacronym{gsfc}{GSFC}{\gls{nasa} Goddard Space Flight Center}
\newacronym{stsci}{STScI}{Space Telescope Science Institute}
\newacronym{nsroc}{NSROC}{\gls{nasa} Sounding Rocket Operations Contract}
\newacronym{wff}{WFF}{\gls{nasa} Wallops Flight Facility}
\newacronym{wsmr}{WSMR}{White Sands Missile Range}

\newacronym{irac}{IRAC}{Infrared Array Camera}
\newacronym[plural=CCDs, firstplural=charge-coupled devices (CCDs)]{ccd}{CCD}{charge-coupled device}
\newacronym[plural=EMCCDs, firstplural=electron multiplying charge-coupled devices (EMCCDs)]{EMCCD}{EMCCD}{electron multiplying charge-coupled device}

\newacronym{DM}{DM}{Deformable Mirror}
\newacronym{MCP}{MCP}{ Microchannel Plate }
\newacronym{ipc}{IPC}{Image Proportional Counter}
\newacronym{cots}{COTS}{Commercial Off-The-Shelf}
\newacronym{ISR}{ISR}{incoherent scatter radar}
\newacronym{atcamera}{AT}{angle tracker}
\newacronym{MEMS}{MEMS}{microelectromechanical systems}
\newacronym{QE}{QE}{quantum efficiency}
\newacronym{RTD}{RTD}{Resistance Temperature Detector}
\newacronym{PID}{PID}{Proportional-Integral-Derivative}
\newacronym{PRNU}{PRNU}{photo response non-uniformity}
\newacronym{DSNU}{PRNU}{dark signal non-uniformity}
\newacronym{CMOS}{CMOS}{complementary metal–oxide–semiconductor}
\newacronym{TRL}{TRL}{technology readiness level}
\newacronym{swap}{SWaP}{Size, Weight, and Power}
\newacronym{ConOps}{ConOps}{concept of operations}
\newacronym{NRE}{NRE}{non-recurring engineering}
\newacronym{CBE}{CBE}{current best estimate}

\newacronym{FOV}{FOV}{field-of-view}
\newacronym{NIR}{NIR}{near-infrared}
\newacronym{PV}{PV}{Peak-to-Valley}
\newacronym{MRF}{MRF}{Magnetorheological finishing}
\newacronym{AO}{AO}{Adaptive Optics}
\newacronym{TTP}{TTP}{tip, tilt, and piston}
\newacronym{FPS}{FPS}{fine pointing system}
\newacronym{SHWFS}{SHWFS}{Shack-Hartmann Wavefront Sensor}
\newacronym{OAP}{OAP}{off-axis parabola}
\newacronym{LGS}{LGS}{laser guide star}
\newacronym{WFCS}{WFCS}{wavefront control system}
\newacronym{OPD}{OPD}{optical path difference}
\newacronym{UA}{UA}{University of Arizona}
\newacronym{MEL}{MEL}{Master Equipment List}
\newacronym{LEO}{LEO}{low-earth orbit}
\newacronym{GEO}{GEO}{geosynchronous orbit}
\newacronym{EFC}{EFC}{electric-field conjugation}
\newacronym{LDFC}{LDFC}{linear dark field control}
\newacronym{DAC}{DAC}{digital-to-analog converter}
\newacronym{FEA}{FEA}{finite element analysis}

\newacronym{SiC}{SiC}{Silicon Carbide}
\newacronym{ESPA}{ESPA}{EELV Secondary Payload Adapter}
\newacronym{EEID}{EEID}{Earth-equivalent Insolation Distance, the distance from the star where the incident energy density is that of the Earth received from the Sun}
\newacronym{LLOWFS}{LLOWFS}{Lyot low-order wavefront sensor}
\newacronym{STOP}{STOP}{Structural-Thermal-Optical-Performance}

\newacronym{resel}{resel}{resolution element}

\newacronym{acs}{ACS}{Attitude Control System}
\newacronym{orsa}{ORSA}{Ogive Recovery System Assembly}
\newacronym{gse}{GSE}{Ground Station Equipment}
\newacronym{FSM}{FSM}{Fast Steering Mirror}

\newacronym{WFS}{WFS}{wavefront sensor}
\newacronym{LSI}{LSI}{Lateral Shearing Interferometer}
\newacronym{VVC}{VVC}{Vector Vortex Coronagraph}
\newacronym{VNC}{VNC}{Visible Nulling Coronagraph}
\newacronym{CGI}{CGI}{Coronagraph Instrument}
\newacronym{IWA}{IWA}{Inner Working Angle}
\newacronym{OWA}{OWA}{Outer Working Angle}
\newacronym{NPZT}{N-PZT}{Nuller piezoelectric transducer}
\newacronym{ZWFS}{ZWFS}{Zernike wavefront sensor}
\newacronym{SPC}{SPC}{Shaped Pupil Coronagraph}
\newacronym{HLC}{HLC}{Hybrid-Lyot Coronagraph}
\newacronym{ADI}{ADI}{angular differential imaging}
\newacronym{RDI}{RDI}{reference differential imaging}
\newacronym{LOWFSC}{LOWFS/C}{low-order wavefront sensing and control}
\newacronym{HOWFSC}{HOWFS/C}{high-order wavefront sensing and control}
\newacronym{WFSC}{WFSC}{wavefront sensing and control}

\newacronym{HST}{HST}{Hubble Space Telescope}
\newacronym{GPS}{GPS}{Global Positioning System}
\newacronym{ISS}{ISS}{International Space Station}
\newacronym[description=Advanced CCD Imaging Spectrometer]{acis}{ACIS}{Advanced \gls{ccd} Imaging Spectrometer}
\newacronym{stis}{STIS}{\textit{Space Telescope Imaging Spectrograph}}
\newacronym{mcp}{MCP}{Microchannel Plate}
\newacronym{JWST}{JWST}{$\textit{James Webb Space Telescope}$}
\newacronym{fuse}{FUSE}{$\textit{FUSE}$}
\newacronym{galex}{GALEX}{$\textit{Galaxy Evolution Explorer}$}
\newacronym{spitzer}{Spitzer}{$\textit{Spitzer Space Telescope}$}
\newacronym{mips}{MIPS}{Multiband Imaging Photometer for \gls{spitzer}}
\newacronym{gissmo}{GISSMO}{Gas Ionization Solar Spectral Monitor}
\newacronym{iue}{IUE}{International Ultraviolet Explorer}
\newacronym{spinr}{SPINR}{$\textit{Spectrograph for Photometric Imaging with Numeric Reconstruction}$}
\newacronym{imager}{IMAGER}{$\textit{Interstellar Medium Absorption Gradient Experiment Rocket}$}
\newacronym{TPF-C}{TPF-C}{Terrestrial Planet Finder Coronagraph}
\newacronym{RAIDS}{RAIDS}{Atmospheric and Ionospheric Detection System }
\newacronym{mama}{MAMA}{Multi-Anode Microchannel Array}
\newacronym{ATLAST}{ATLAST}{Advanced Technology Large Aperture Space Telescope}
\newacronym{PICTURE}{PICTURE}{Planet Imaging Concept Testbed Using a Rocket Experiment}
\newacronym{LITES}{LITES}{Limb-imaging Ionospheric and Thermospheric
Extreme-ultraviolet Spectrograph}
\newacronym{LBT}{LBT}{Large Binocular Telescope}
\newacronym{LBTI}{LBTI}{Large Binocular Telescope Interferometer}
\newacronym{KIN}{KIN}{Keck Interferometer Nuller}
\newacronym{SHARPI}{SHARPI}{Solar High-Angular Resolution Photometric Imager}
\newacronym{IRAS}{IRAS}{Infrared Astronomical Satellite}
\newacronym{HARPS}{HARPS}{High Accuracy Radial velocity Planetary}
\newacronym{hstSTIS}{STIS}{Space Telescope Imaging Spectrograph}
\newacronym{spitzerIRAC}{IRAC}{Infrared Array Camera}
\newacronym{spitzerMIPS}{MIPS}{Multiband Imaging Photometer for Spitzer}
\newacronym{spitzerIRS}{IRS}{Infrared Spectrograph}
\newacronym{CHARA}{CHARA}{Center for High Angular Resolution Astronomy}
\newacronym{wfirst-afta}{WFIRST-AFTA}{Wide-Field InfrarRed Survey
Telescope-Astrophysics Focused Telescope Assets}
\newacronym{GPI}{GPI}{Gemini Planet Imager}
\newacronym{Roman}{Roman}{Nancy Grace Roman Space Telescope}
\newacronym{HabEx}{HabEx}{Habitable Exoplanet Observatory Mission Concept}
\newacronym{LUVOIR}{LUVOIR}{Large UV/Optical/Infrared Surveyor}
\newacronym{FGS}{FGS}{Fine Guidance Sensor}
\newacronym{STIS}{STIS}{Space Telescope Imaging Spectrograph}
\newacronym{MGHPCC}{MGHPCC}{Massachusetts Green High Performance
Computing Center}
\newacronym{WISE}{WISE}{Wide-field Infrared Survey Explorer}
\newacronym{ALMA}{ALMA}{Atacama Large Millimeter Array}
\newacronym{GRAIL}{GRAIL}{Gravity Recovery and Interior Laboratory}
\newacronym{jwstNIRCam}{NIRCam}{Near-Infrared Camera}
\newacronym{jwstMIRI}{MIRI}{Mid-Infrared Instrument}

\newacronym{AURIC}{AURIC}{The Atmospheric Ultraviolet Radiance Integrated Code} 
\newacronym{FFT}{FFT}{Fast Fourier Transform  }
\newacronym{MODTRAN}{MODTRAN   }{ MODerate resolution atmospheric TRANsmission }
\newacronym{idl}{IDL}{$\textit {Interactive Data Language}$}
\newacronym[sort=NED,description=NASA/IPAC Extragalactic Database]{ned}{NED}{\gls{nasa}/\gls{ipac} Extragalactic Database}
\newacronym{iraf}{IRAF}{Image Reduction and Analysis Facility}
\newacronym{wcs}{WCS}{World Coordinate System}
\newacronym{pegase}{PEGASE}{$\textit{Projet d'Etude des GAlaxies par Synthese Evolutive}$}
\newacronym{dirty}{DIRTY}{$\textit{DustI Radiative Transfer, Yeah!}$}
\newacronym{CUDA}{CUDA}{Compute Unified Device Architecture}
\newacronym{KLIP}{KLIP}{Karhunen-Lo\`eve Image Processing}

\newacronym{MSIS}{MSIS}{Mass Spectrometer Incoherent Scatter Radar}
\newacronym{nmf2}{$N_m$}{F2-Region Peak density}
\newacronym{hmf2}{$h_m$}{F2-Region Peak height}
\newacronym{H}{$H$}{F2-Region Scale Height}

\newacronym{isr}{ISR}{Incoherent Scatter Radar}
\newacronym[description=TLA Within Another Acronym]{twaa}{TWAA}{\gls{tla} Within Another Acronym}
\newacronym[plural=SNe, firstplural=Supernovae (SNe)]{sn}{SN}{Supernova}
\newacronym{EUV}{EUV}{Extreme-Ultraviolet }
\newacronym{EUVS}{EUVS}{\gls{EUV} Spectrograph}
\newacronym{F2}{F2}{Ionospheric Chapman F Layer }
\newacronym{F10.7}{F10.7}{ 10.7 cm radio flux [10$^{-22}$ W m$^{-2}$ Hz$^{-1}$]  }
\newacronym{FUV}{FUV}{far-ultraviolet }
\newacronym{IR}{IR}{infrared}
\newacronym{MUV}{MUV}{mid-ultraviolet }
\newacronym{NUV}{NUV}{near-ultraviolet }
\newacronym{O$^+$}{O$^+$}{Singly Ionized Oxygen  Atom }
\newacronym{OI}{OI}{Neutral Atomic Oxygen Spectroscopic State }
\newacronym{OII}{OII}{Singly Ionized Atomic Oxygen Spectroscopic State }
\newacronym{PSF}{PSF}{point spread function}
\newacronym{$R_E$}{$R_E$}{Earth radii [$\approx$ 6400 km]  }
\newacronym{RV}{RV}{radial velocity}
\newacronym{UV}{UV}{ultraviolet }
\newacronym{WFE}{WFE}{wavefront error}
\newacronym{sed}{SED}{spectral energy distribution}
\newacronym{nir}{NIR}{near-infrared}
\newacronym{mir}{MIR}{mid-infrared}
\newacronym{ir}{IR}{infrared}
\newacronym{uv}{UV}{ultraviolet}
\newacronym[plural=PAHs, firstplural=Polycyclic Aromatic Hydrocarbons (PAHs)]{pah}{PAH}{Polycyclic Aromatic Hydrocarbon}
\newacronym{obsid}{OBSID}{Observation Identification}
\newacronym{SZA}{SZA}{Solar Zenith Angle}
\newacronym{TLE}{TLE}{Two Line Element set}
\newacronym{DOF}{DOF}{degrees-of-freedom}
\newacronym{PZT}{PZT}{lead zirconate titanate}
\newacronym{ADCS}{ADCS}{attitude determination and control system}
\newacronym{COTS}{COTS}{commercial off-the-shelf}
\newacronym{CDH}{C$\&$DH}{command and data handling}
\newacronym{EPS}{EPS}{electrical power system}

\newacronym{PCA}{PCA}{principal component analysis}
\newacronym{fwhm}{FWHM}{full-width-half maximum}
\newacronym{RMS}{RMS}{root mean squared}
\newacronym{RMSE}{RMSE}{root mean squared error}
\newacronym{MCMC}{MCMC}{Marcov chain Monte Carlo}
\newacronym{DIT}{DIT}{discrete inverse theory}
\newacronym{SNR}{SNR}{signal-to-noise ratio}
\newacronym{PSD}{PSD}{power spectral density}
\newacronym{NMF}{NMF}{Non-negative Matrix Factorization}

\begin{document}
\begin{CJK*}{UTF8}{gbsn}

\title{Deepest limits on scattered light emission from the Epsilon Eridani inner debris disk with HST/STIS}

\correspondingauthor{Sai Krishanth P.M.}
\email{saikrishanth@arizona.edu}

\author[0000-0002-9869-6223]{Sai Krishanth P.M.}
\affiliation{Steward Observatory,  The University of Arizona, 933 North Cherry Avenue, Tucson, AZ 85721, USA}

\author[0000-0002-0813-4308]{Ewan S. Douglas}
\affiliation{Steward Observatory,  The University of Arizona, 933 North Cherry Avenue, Tucson, AZ 85721, USA}

\author[0000-0002-4989-6253]{Ramya M. Anche}
\affiliation{Steward Observatory,  The University of Arizona, 933 North Cherry Avenue, Tucson, AZ 85721, USA}

\author[0000-0001-9994-2142]{Justin Hom}
\affiliation{Steward Observatory,  The University of Arizona, 933 North Cherry Avenue, Tucson, AZ 85721, USA}

\author[0000-0002-7791-5124]{Kerri L. Cahoy}
\affiliation{Department of Aeronautics and Astronautics, Massachusetts Institute of Technology, Cambridge, MA 02139, USA}
\affiliation{Department of Earth, Atmospheric, and Planetary Sciences, Massachusetts Institute of Technology, Cambridge, MA 02139, USA}

\author[0000-0002-1783-8817]{John H. Debes}
\affiliation{Space Telescope Science Institute, 3700 San Martin Drive, Baltimore, MD 21218, USA}

\author[0000-0002-7639-1322]{Hannah Jang-Condell}
\affiliation{Department of Physics \& Astronomy, University of Wyoming, Laramie, WY 82071, USA}
\affiliation{NASA Headquarters, Washington, DC 20546}
\author[0000-0002-4989-6253]{Isabel Rebollido}
\affiliation{European Space Agency (ESA), European Space Astronomy Centre (ESAC), Camino Bajo del Castillo s/n, 28692, Villanueva de la Ca\~nada, Madrid, Spain}

\author[0000-0003-1698-9696]{Bin B. Ren (任彬)}
\altaffiliation{Marie Sk\l odowska-Curie Fellow}
\affiliation{Universit\'{e} C\^{o}te d'Azur, Observatoire de la C\^{o}te d'Azur, CNRS, Laboratoire Lagrange, Bd de l'Observatoire, CS 34229, 06304 Nice cedex 4, France}
\affiliation{Universit\'{e} Grenoble Alpes, CNRS, Institut de Plan\'{e}tologie et d'Astrophysique (IPAG), F-38000 Grenoble, France}
\affiliation{Department of Astronomy, California Institute of Technology,  MC 249-17,1200 East California Boulevard, Pasadena, CA 91125, USA}

\author{Christopher C. Stark}
\affiliation{NASA Goddard Space Flight Center, Greenbelt, MD 20771, USA}

\author{Robert Thompson}
\affiliation{Steward Observatory,  The University of Arizona, 933 North Cherry Avenue, Tucson, AZ 85721, USA}

\author[0000-0002-6171-9081]{Yinzi Xin}
\affiliation{Department of Aeronautics and Astronautics, Massachusetts Institute of Technology, Cambridge, MA 02139, USA}


\begin{abstract}
\gls{epseri} is one of the first debris disk systems detected by the \gls{IRAS}. However, the system has thus far eluded detection in scattered light with no components having been directly imaged. Its similarity to a relatively young Solar System combined with its proximity makes it an excellent candidate to further our understanding of planetary system evolution. We present a set of coronagraphic images taken using the \gls{STIS} coronagraph on the Hubble space telescope at a small inner working angle to detect a predicted warm inner debris disk inside 1\arcsec. We used three different post-processing approaches; \gls{NMF}, \gls{KLIP}, and Classical \gls{RDI}, to best optimize reference star subtraction, and find that \gls{NMF} performed the best overall while \gls{KLIP} produced the absolute best contrast inside 1\arcsec. 
We present limits on scattered light from warm dust, with constraints on surface brightness at 6 mJy/as$^2$ at our inner working angle of 0.6\arcsec. We also place a constraint of 0.5 mJy/as$^2$ outside 1\arcsec, which gives us an upper limit on the brightness for outer disks and substellar companions. Finally, we calculated an upper limit on the dust albedo at $\omega$ $<$ 0.487.  
\end{abstract}


\section{Introduction}\label{intro}
\Gls{epseri} (K2V; 3.2 pc; \citealt{gaiadr2}) is one of the nearest solar analogs and presents an ideal laboratory for testing our understanding of debris disk formation and evolution.
\label{sec:eps_eri_dusty_lab}
 \gls{epseri} was one of the earliest extrasolar debris disks observed, discovered by the \gls{IRAS} detection of emission in exceptional excess of what would be expected from the stellar photosphere \citep{aumann_iras_1985}.
\cite{greaves_dust_1998} first resolved a cold outer disk morphology at 450 and 850 $\mu$m, finding a cool, nearly face-on \gls{EKB} analog from 30-75 AU. This was shown to be a uniform, circular ring using the Submillimeter Array \citep{macgregor_epsilon_2015}. 
 
 \gls{epseri}'s \gls{IR}-excess at 24 $\mu$m \citep{backman_epsilon_2009} is two orders of magnitude brighter than the solar zodiacal dust  \citep{backman_main-sequence_1993, backman_epsilon_2009} and may present a significant background signal which must be overcome by future direct imaging missions to image the nearest known giant exoplanet, \gls{epseri} b \citep{hatzes_evidence_2000}.

 A wide range of orbital parameters have been reported in the literature, and as observations have accumulated, the presence of this companion has become more clear. 
\cite{howard_limits_2016} reported a ``clear detection" with an eccentricity of 0.26. 
\cite{mawet_deep_2018} performed a joint analysis of both direct imaging non-detection sensitivity and radial velocity observations, constraining the companion parameters to a mass of $0.78^{+0.38}_{-0.12}$ $M_{Jup}$ with a semi-major axis of $3.48\pm 0.02$ au and a nearly circular orbit (e = $0.07^{+0.06}_{-0.05}$).
\cite{hunziker_refplanets_2019} observed \gls{epseri} with the SPHERE ZIMPOL instrument with the VBB filter (broadband Vis-NIR filter at 735 nm). 
By combining polarimetric and \gls{ADI}, they placed  $1\sigma$ limits on the polarized intensity approaching contrasts of $10^{-8}$ at 1\arcsec\  separation and 0.5-1e-7 sensitivity to un-polarized point sources at 1\arcsec. 
\cite{llop-sayson_constraining_2021} applied ground-based direct imaging non-detection limits, \gls{RV}, and astrometric constraints to find a lower mass $0.66^{+.12}_{-.09}$ and a relatively high inclination $i=78.81^{+29.34}_{-22.41}$. 
\cite{benedict_revisiting_2022} re-analyzed Hubble \gls{FGS} astrometry observations \citep{benedict_extrasolar_2007} in the context of these new observations and found $i=45\pm8$. 
Additionally, the 42.7$_{9.5}^{11}$ degree obliquity between the resolved outer debris disk(s) and the star \citep{hurt_evidence_2023} suggests \gls{epseri} may have a formation history significantly different from the Solar System and that the inner disk's inclination is far from certain. 

Scattered light imaging efforts to resolve questions about the location, composition, and orientation of the debris around \gls{epseri} have proven unsuccessful.
A search by \cite{proffitt_limits_2004}  with the \gls{hstSTIS} put a limit of 0.378 $\mu$Jy/arcsecond$^{2}$ for scattered light from the outer ring. And most recently, \cite{wolff2023}, using \gls{hstSTIS}, placed a constraint of 4 $\mu$Jy/arcsecond$^{2}$ at radial separations outside 10\arcsec. 

Combining new observations with \gls{NMF} \gls{PSF} subtraction, this work will present new limits on the broadband un-polarized scattered light from extended and point sources around \gls{epseri} using the \gls{STIS} instrument aboard \gls{HST}. In Section \ref{observations}, we describe the observation program used to collect data using the WEDGE A 1.0 occulter position, and in Section \ref{reduction}, we present our reduction strategy, with NMF being our primary method of post-processing. In Section \ref{results}, we summarize our results, including a new contrast floor, and describe the PSF color mismatch problem that affects classical RDI subtraction with \gls{STIS}. Finally, in Section \ref{discussion}, we discuss future observatories and their ability to directly image the \gls{epseri} inner disk in scattered light. We also compare our different post-processing methods and present some extensions of our NMF reduction. 

\section{Observations}\label{observations}

We observed \gls{epseri} with the \gls{HST} \gls{STIS} WEDGE A 1.0 coronagraph \citep{grady_coronagraphic_2003} using the 50CORON aperture (central wavelength of 5743.706 \AA \ with a wavelength range from 1640 \AA \ to 10,300 \AA, \gls{fwhm} = 4333 \AA) with 1024$\times$110 pixel frames to minimize readout time and a gain of 4 e$^-$/ADU. Our observations were part of \gls{HST} cycle 25 observations (PID 15217).
Exposure times of 2.3 seconds were set to prevent saturation at the inner working angle by comparison with other STIS coronagraph observations.
Observations were carried out in two campaigns, the first visit (V-I) on 2018-11-28 starting at 06:53:59 UTC and a second visit (V-II) on 2018-12-27 starting at 16:28:59 UTC.
V-I and V-II included three orbits of \gls{epseri}, each at a different roll angle separated by 15 degrees, and one orbit of reference star \gls{deleri}. A summary of these observations is shown in Table \ref{table:obs_table}. 

\gls{deleri}, a K0+IV  star 9 pc distant, was chosen because of excellent color matching (the B-V color differs from \gls{epseri} by only 0.04 mag), proximity on the sky minimizing thermal changes, and lack of known circumstellar material.
\gls{deleri} exposure times were decreased to 1.9 sec to provide comparable counts per frame; This is done in part to mitigate differences in PSF shape due to charge transfer inefficiency. Even at high count rates where little residual charge is lost, differences at the 10$^{-8}$ contrast level are possible if the count rates on the STIS CCD differ by more than a factor of a few \citep{debes_pushing_2019}.
V-I was designed to maximize \gls{PSF} library matching using sub-pixel dithering, with 3 pointings across the mask separated 0.25 pix, which has previously shown promising results with the BAR5 coronagraph \citep{debes_pushing_2019}. 

In October 2018, Gyro 2 on HST failed and was replaced with Gyro 3. Gyro 3, with a higher level of rate bias shifts than previous HST Gyros, made target acquisitions fail at a higher than historical rate. For visit V-I, the Fine Guidance Sensor was only able to acquire on a single guide star, causing a slow drift to the stellar pointing over the course of each visit and enhancing the overall level of jitter (average $\sigma$=10.5~mas over an orbit). The drift itself was not sufficient to render the observations unusable but decreased the similarity between observations and the past observations used to generate the reference library. HST pointing jitter, in particular, can impact contrast at small inner working angles \citep{debes_pushing_2019}. 

For V-II, subpixel dithering was removed from the program and three additional roll angles were observed along with a second observation of \gls{deleri}. The target acquisition for these visits was nominal, and the jitter level, as measured by the stellar centroids was slightly improved over V-I (average $\sigma$=9~mas over an orbit).  
The total roll angle space spanned was 75 degrees, from 176.33$^{\circ}$ to 251.33$^{\circ}$ in \gls{HST} V-3 roll coordinates. The total integration time was 1507 seconds on \gls{epseri} and 455 seconds on $\delta$-Eri.

\begin{table*}[h]
\begin{centering}
\begin{tabular}{@{}llllllll@{}}
\toprule
Visit & Target    & Start Time (UT)      & Roll Angle (deg) & Subpixel dithering (pix) & Exposure time (s) & N$_{\text{images}}$ & GS Acquisition\\ \midrule \cmidrule(r){1-7}
V-I-4  & \gls{epseri}         & 2018-11-28 06:58:49 & 206.33 & -0.25 & 83.99      & 31 & GSFAIL         \\ 
V-I-4  & \gls{epseri}         & 2018-11-28 07:13:15 & 206.33 &  0    & 98.29      & 31 & GSFAIL         \\
V-I-4  & \gls{epseri}         & 2018-11-28 07:27:55 & 206.33 & +0.25 & 83.99      & 31 & GSFAIL         \\
V-I-5  & \gls{epseri}         & 2018-11-28 08:34:10 & 191.33 & -0.25 & 83.99      & 31 & GSFAIL         \\
V-I-5  & \gls{epseri}         & 2018-11-28 08:48:36 & 191.33 & 0     & 98.29      & 31 & GSFAIL         \\
V-I-5  & \gls{epseri}         & 2018-11-28 09:03:16 & 191.33 & +0.25 & 83.99      & 31 & GSFAIL         \\
V-I-8  & \gls{deleri} (PSF)   & 2018-11-28 10:10:09 & 200.89 & -0.25 & 71.30      & 32 & OK             \\
V-I-8  & \gls{deleri} (PSF)   & 2018-11-28 10:24:28 & 200.89 &  0    & 83.70      & 32 & OK             \\
V-I-8  & \gls{deleri} (PSF)   & 2018-11-28 10:38:58 & 200.89 & +0.25 & 71.30      & 32 & OK             \\
V-I-6  & \gls{epseri}         & 2018-11-28 11:44:51 & 176.33 & -0.25 & 83.99      & 31 & GSFAIL         \\
V-I-6  & \gls{epseri}         & 2018-11-28 11:59:17 & 176.33 &  0    & 98.30      & 31 & GSFAIL         \\
V-I-6  & \gls{epseri}         & 2018-11-28 12:13:57 & 176.33 & +0.25 & 83.99      & 31 & GSFAIL         \\ 
 \hline
V-II-1 & \gls{epseri}         & 2018-12-27 16:33:53 & 221.33 & 0     & 233.50     & 95 & OK             \\
V-II-2 & \gls{epseri}         & 2018-12-27 19:44:35 & 236.33 & 0     & 233.50     & 95 & OK             \\
V-II-3 & \gls{epseri}         & 2018-12-27 18:09:59 & 251.33 & 0     & 233.50     & 95 & OK             \\
V-II-7 & \gls{deleri} (PSF)   & 2018-12-27 21:19:57 & 229.02 & 0     & 200.99     & 99 & OK             \\ \bottomrule
\end{tabular}
\caption{Summary of observations made with our GO 15217 program. GSFAIL indicates observations where the \gls{FGS} was unable to lock on to more than one guide star, leading to a higher rate of pointing drift than nominal.}
\label{table:obs_table}
\end{centering}
\end{table*}

 \begin{figure*}[h]
    \centering
    \includegraphics[width=0.95\textwidth]{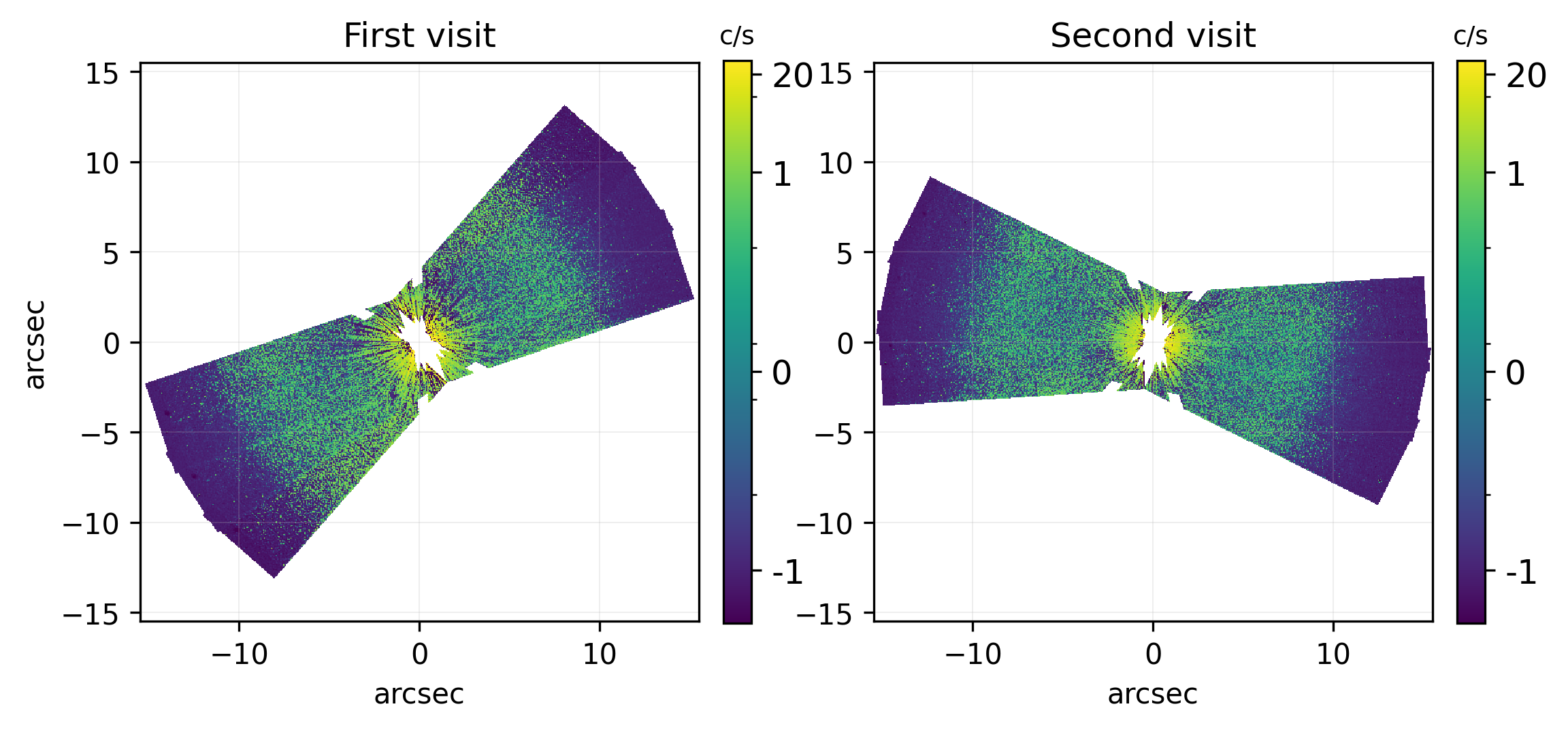}
    \caption{Median combined images for each visit. The mean of the median of each roll for the first and last visit shows relatively similar structure and brightness between the two visits. No images were dropped before the median combining shown in this illustration.}
    \label{fig:combined_visits}
\end{figure*}

\section{Reduction} \label{reduction}

\subsection{Pre-processing}

We obtained raw and uncalibrated \gls{STIS} WEDGE A 1.0 images from the MAST\footnote{\url{https://mast.stsci.edu/}} archive. Following \cite{wolff2023}, we used the AUTOFILET package \citep{autofilet} to remove the readout video noise present in all CCD side 2 observations since 2001. The video noise-removed raw frames were processed through the standard calstis pipeline\footnote{\url{https://stistools.readthedocs.io/en/latest/calstis.html}}, yielding dark and bias-subtracted flat-fielded frames. Coronagraphic flats were not needed as the flux error due to differences in the flat field across the image is less than 2\% and impacts both the target and reference stars equally. We constructed an algorithmic masking function for the occulting wedge of the coronagraph and the diffraction spikes produced by it. The central wedge mask had a \gls{fwhm} of 1.1\arcsec while the masked diffraction spikes had a thickness of 0.7\arcsec. We then checked every frame by eye to ensure mask efficacy, ensuring minimal loss of scientific information. We find significant improvement in contrast performance with well-masked frames, with higher recovery of data behind the mask, and prevention of overfitting during post-processing.
Our data reduction from this point follows the STIS coronagraph pipeline described in \cite{ren_post-processing_2017}. 

\subsection{Post-processing}

We undertook three post-processing approaches, classical \gls{PSF} subtraction, \gls{KLIP} \citep{soummer_detection_2012,amara_pynpoint_2012}, and \gls{NMF} \citep{ren_non-negative_2018}. A library of \gls{STIS} WEDGE A 1.0 images of \gls{deleri} was used to produce a \gls{RDI} \gls{PSF} library for all three post-processing approaches applied, totaling $380$ individual reference readout images.

\subsubsection{NMF}

\begin{figure*}[h!]
    \centering
    \includegraphics[width=0.95\textwidth]{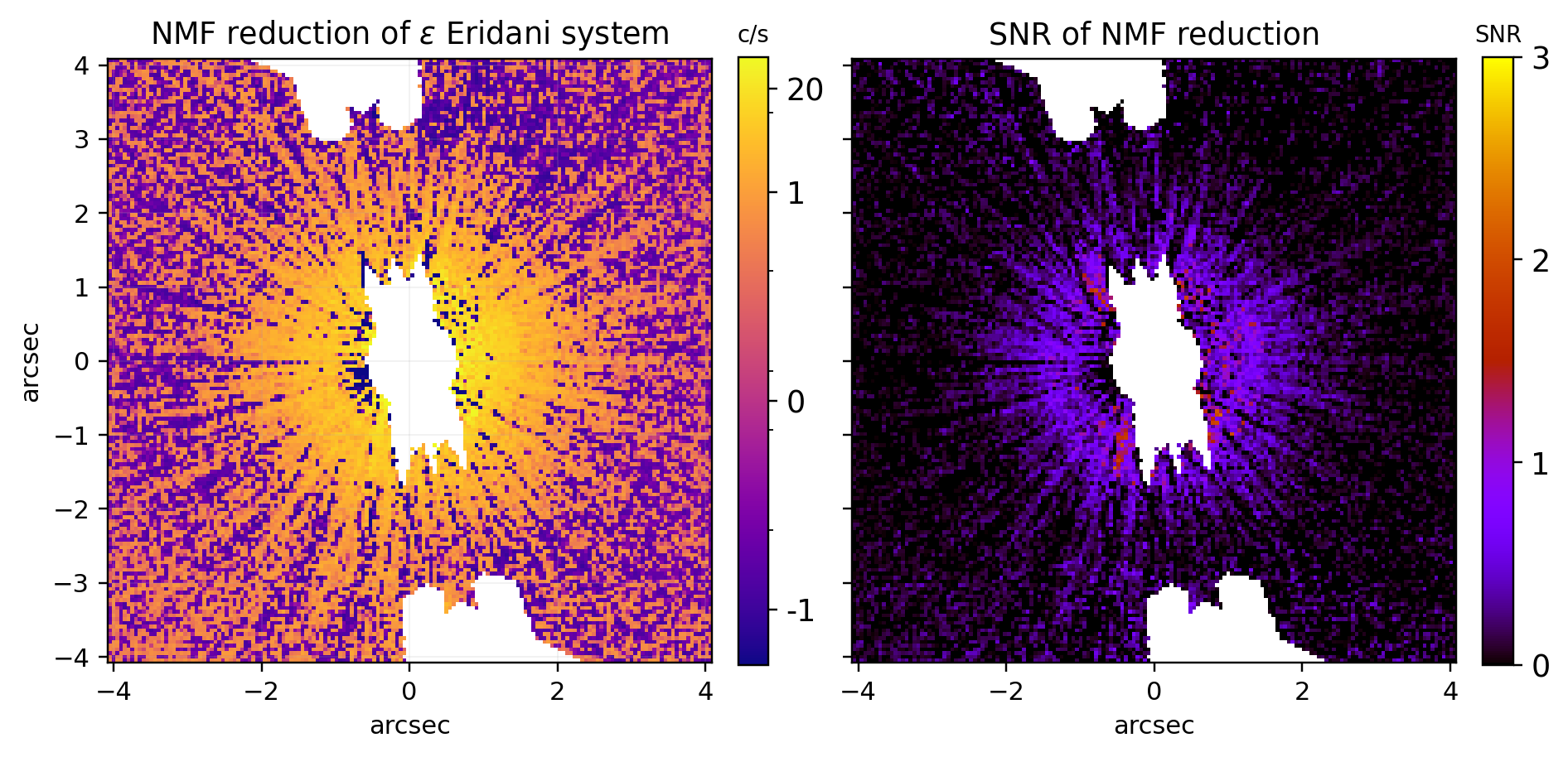}
    \caption{Left panel: NMF result using 90\% closest frames of  \gls{deleri} as reference and 30 NMF components. Right panel: \gls{SNR} map of the reduced image, calculated by dividing the median of used frames by the standard deviation of the used frames.}
    \label{fig:nmf_final}
\end{figure*}

Our NMF implementation is described in detail in \cite{ren_non-negative_2018} and \citet{skpmgpunmf2023}. We use $90\%$ of the nearest reference images in correlation -- 342 frames -- to model the target images with $30$ \gls{NMF} components. By inspection, we found a minor increase in contrast when using 30 components over 10, but larger numbers of components did not yield a significant improvement in contrast. After NMF-mode subtraction, we applied 3$\sigma$ clipping around the data median to every frame. The frames were then rotated into sky coordinates (north up, east left) and cropped to 600$\times$110 pixels. Due to the combination of pointing offsets and the lack of \gls{FGS} lock, some frames were significantly off the coronagraphic wedge and were poorly subtracted. These frames were removed by sorting the post-subtraction frames by the standard deviation of the cropped residual images and discarding the 10\% of frames with the highest standard deviation. The remaining frames were median combined to produce a final image, as shown in Figure \ref{fig:nmf_final}.

\subsubsection{KLIP}

\begin{figure*}[h!]
    \centering
    \includegraphics[width=0.95\textwidth]{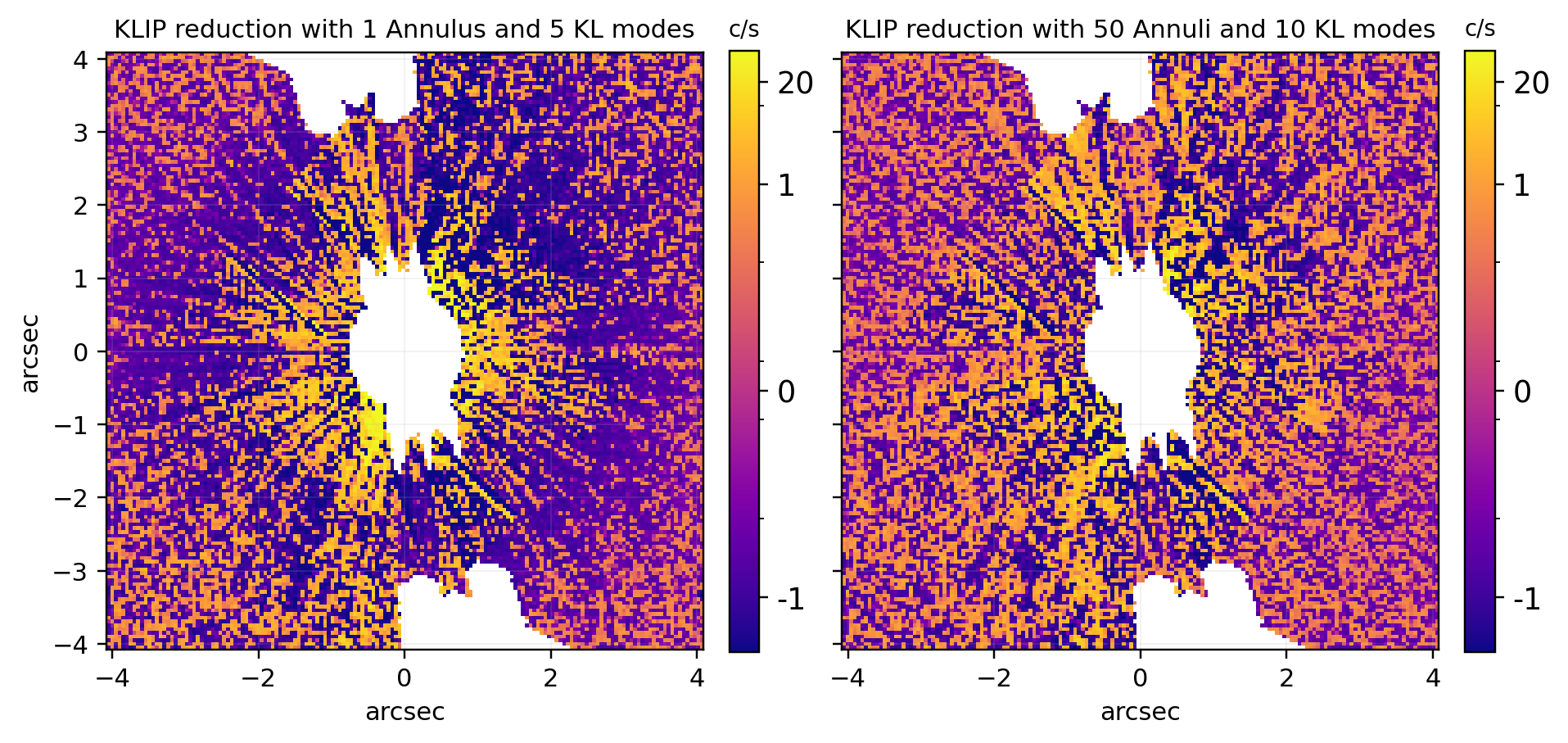}
    \caption{Left panel: Disk-optimized KLIP reduction. The combination of a single annulus and a low number of KL modes is ideal to avoid over-subtraction and preserve the disk signal. Right panel: KLIP reduction optimized for detection of substellar companions. This combination of a large number of annuli gives us the most contrast while 10 KL modes give us a good balance of contrast and throughput.}
    \label{fig:klip}
\end{figure*}

To assess the sensitivity of our observation to substellar companions, we perform PSF subtraction using Karhunen-Lo{\`e}ve Image Projection \citep{soummer_detection_2012} with \texttt{pyKLIP} \citep{wang_2015_pyklip} using a large number of logarithmically-spaced concentric annular search zones further divided into four subsections and a range of Karhunen-Lo{\`e}ve (KL) modes up to 100. Having ``aggressive" parameters such as larger numbers of annuli and KL modes will lead to higher contrast overall. This is ideal for detecting point sources such as companions but leads to significant over-subtraction of extended structures such as disks \citep{soummer_detection_2012}. We preferentially order frames in the \gls{RDI} \gls{PSF} library by calculating a correlation matrix of all science frames and reference frames, similar to KLIP-RDI approaches performed in \cite{duchene_2020_gemini} and \cite{chen_2020_multiband}. Further calibration of the contrast curves measured from the reduced KLIP images is performed by injecting fiducial point sources with known fluxes into the pre-processed dataset at varying separations and position angles. We then perform a KLIP-RDI reduction with the same parameters as the initial reduction. The fluxes from the injected planets are retrieved to assess the throughput of the KLIP reduction, which is then used to calibrate the initial contrast measurement. After exploring the parameter space of different numbers of annuli and KL modes, we chose a combination of 50 annuli and 10 KL modes to optimize our KLIP reduction for substellar companions. We also performed a disk-optimized \gls{KLIP} reduction using 1 annulus and 5 KL modes. We use a single annulus for more continuous flux profiles in extended sources, and 5 KL modes to avoid significant over-subtraction and preserve disk throughput at the expense of lower contrast. The final median combined images of both these reductions are shown in Figure \ref{fig:klip}.

\subsubsection{Classical RDI} \label{clasrdi}
\begin{figure*}
    \centering
    \includegraphics[width=0.55\textwidth]{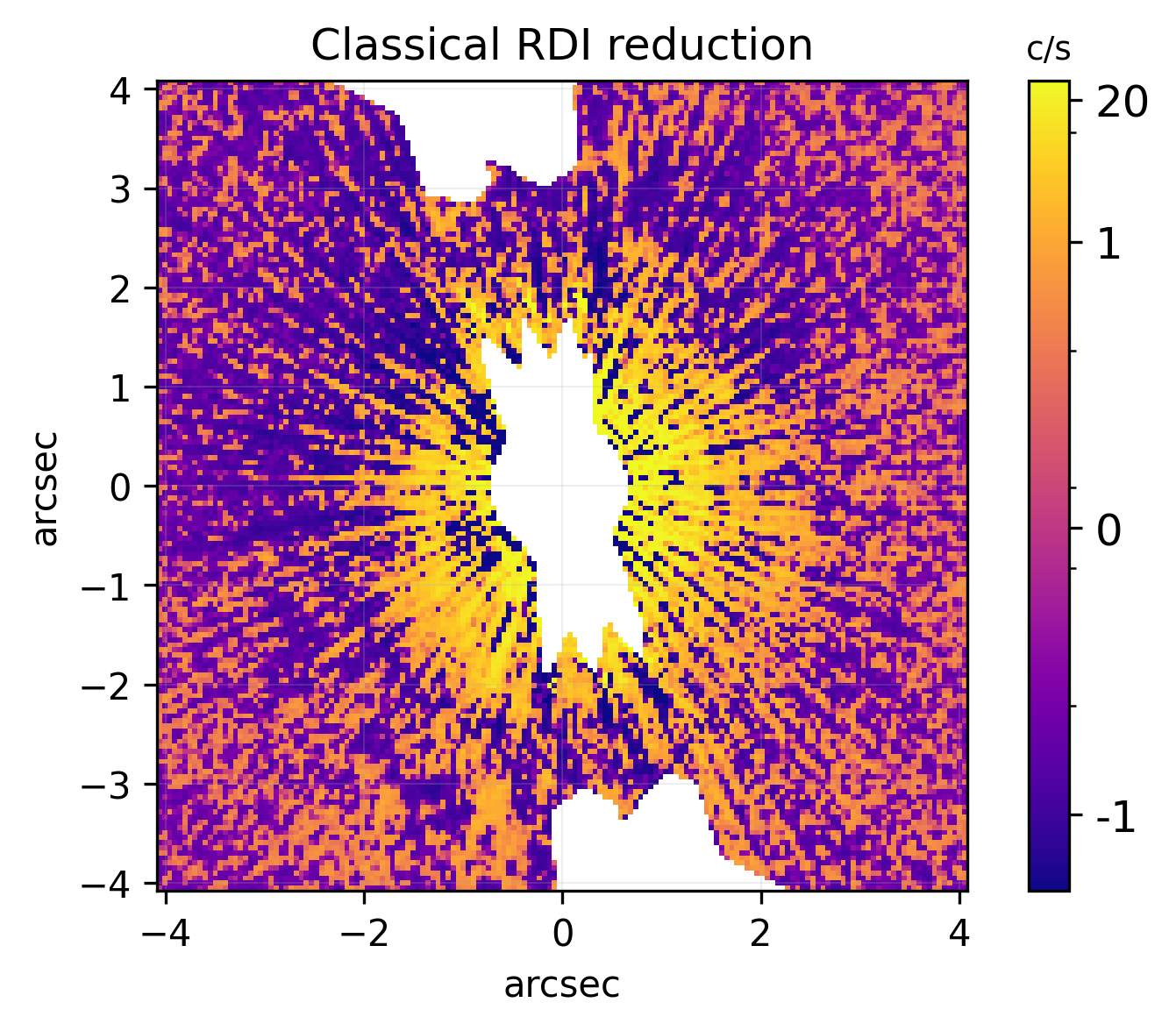}
    \caption{Classical RDI subtraction as described in \ref{clasrdi}.}
    \label{fig:clas_rdi_red}
\end{figure*}

We conducted a classical RDI reduction of the data using the two orbits targeting \gls{deleri} as the sole reference star and largely following the same procedure detailed in \cite{schneider_probing_2014} and \cite{debes_pushing_2019}. Each frame estimated the stellar centroid behind the mask by calculating the intersection of the diffraction spikes, and through subpixel interpolation rectified to a common center \citep{schneider_probing_2014}. 
Differential offsets and scalings between the PSF reference and the target were performed for each orbit, using a mask to occult pixels impacted by the wedge and diffraction spike residuals.
The individual orbits were rotated to the proper position angle on the sky and combined into a final subtracted image. 
This procedure was done for both short exposure times and long exposures at the end of each visit. Both the long and short exposure subtractions were combined into a final image presented in Figure \ref{fig:clas_rdi_red}.

\section{Results}\label{results}

\subsection{Evaluating contrast}\label{evalcont}

To convert to relative contrast values, we used the HST Exposure Time Calculator\footnote{\url{http://etc.stsci.edu/etc/input/stis/imaging/}} to find the global source electron rate (914,424,193 [e/s]) for a V=3.72 mag  K2V star \citep{pickles_stellar_models}. We used this value instead of the ``peak pixel" value to account for the quantum yield correction at shorter wavelengths. The electron rate is converted to counts assuming the nominal gain of 4.087 electrons/DN, for a DN rate of 2.24E8 [ct/s]. Following \cite{wolff2023}, we chose to calculate contrast using \gls{MAD} over standard deviation. As noted by \cite{wolff2023}, \gls{MAD} is a better estimator for smaller data sets, being more robust to outliers. We determined the \gls{MAD} value in 1px wide annuli centered around \gls{epseri} using the \textit{radial\_profile} function in \texttt{poppy} \citep{poppy}. We divided our processed images by the peak pixel value to calculate image sensitivity in contrast units. 
To convert the sensitivity to surface brightness (in mJy/arcsec$^2$), we adopt the conversation factor for a spectrally flat source of  0.1765 mJy/arcsec$^2$ per count per second from previous studies \cite[Table 4]{schneider_probing_2014}. We present contrast curves in Figure \ref{fig:klip_contrast_curves} and Figure \ref{fig:classical_contrast_curves}. In Appendix \ref{sec:appendB}, we further investigate the residuals remaining after \gls{PSF} subtraction to quantify the upper limit on detected flux. Consistent with Figure \ref{fig:nmf_final}, we find that our measured flux residuals are consistent with zero at all separations.

\subsection{Upper Limits on Disk Brightness}

While we were not able to resolve the inner disk, we can place constraints on the surface brightness with the deepest observed contrast floor in scattered light. At the inner working angle of 0.58\arcsec, we note an upper limit on the surface brightness of 5.98 mJy/arcsec$^2$. Additionally, outside 1\arcsec, we place an upper limit of 0.57 mJy/arcsec$^2$. Since the inner disk is expected to be at or under 1\arcsec in diameter \citep{backman_epsilon_2009, su_inner_2017}, this allows us to place an upper limit on the surface brightness of the inner disk. We note that the upper bound placed on disk flux in this work is lower than similar limits established in \cite{wolff2023}, particularly while employing NMF. The improvement in contrast is attributable to an increase in the number of components used during NMF, and to a larger extent uniform sigma clipping across all post processing methods.
In Appendix \ref{appendix:planet}, we also place upper limits on planetary mass using the Sonora-Bobcat models \citep{sonora_bobcat1}. 

\begin{figure*}
    \centering
    \includegraphics[width=0.95\textwidth]{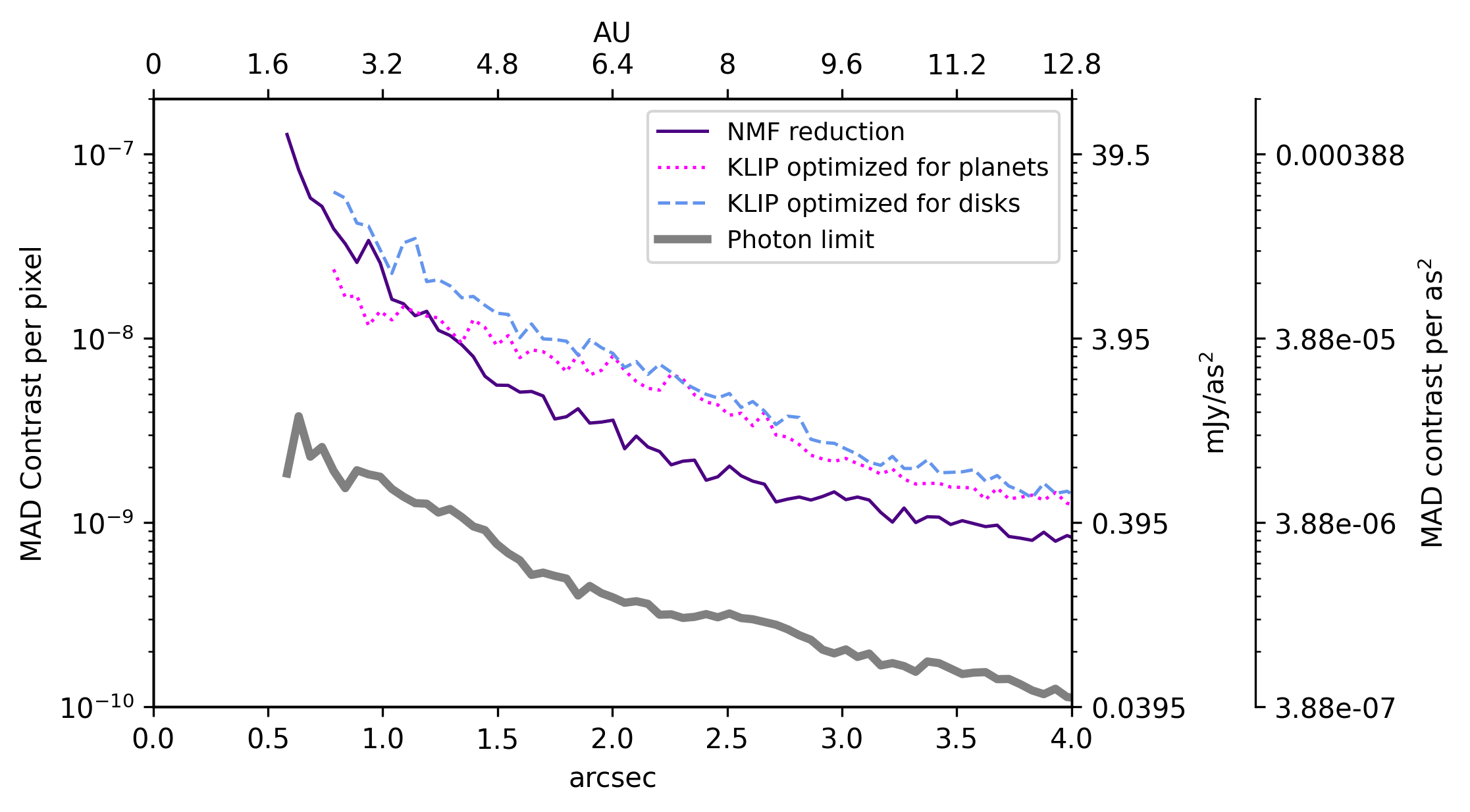}
    \caption{Contrast versus separation sensitivity curves showing the KLIP reduction optimized for disks (in dot-dash blue), KLIP reduction optimized for planets (in dotted magenta), NMF reduction (in solid purple), and the photon limit (in thick solid grey). We note that the KLIP subtraction optimized for planets has the best contrast under 1\arcsec, but the NMF curve has the best contrast overall.}
    \label{fig:klip_contrast_curves}
\end{figure*}

\begin{figure*}
    \centering
    \includegraphics[width=0.95\textwidth]{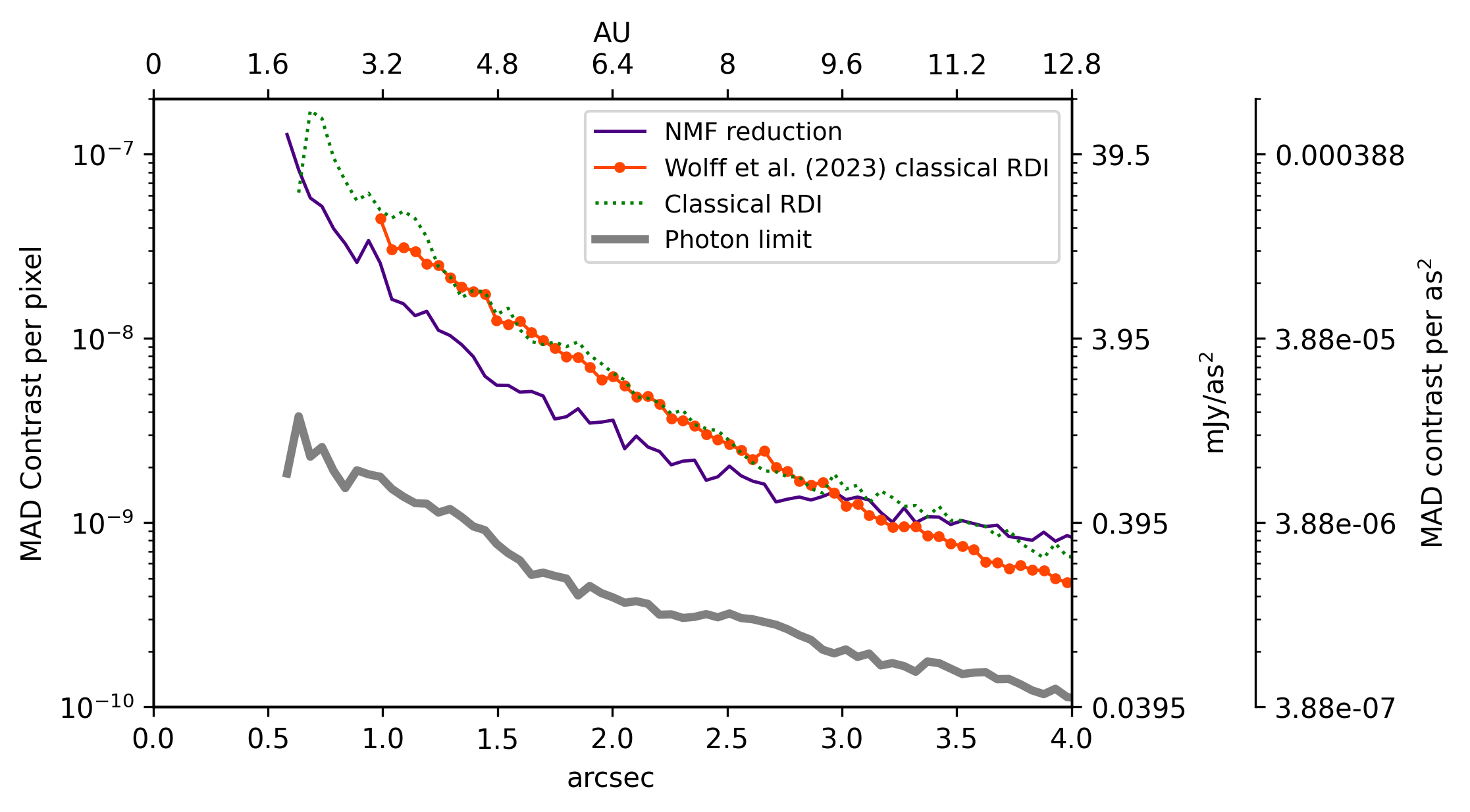}
    \caption{Contrast versus separation sensitivity curves showing our classical \gls{RDI} (in dotted green), classical \gls{RDI} reduction from \cite{wolff2023} (in dot-dash orange), NMF reduction (in solid purple), and the photon limit (in thick solid grey). We note that the NMF curve has better contrast out to 3\arcsec, after which both classical \gls{RDI} reductions perform better. Our reductions also extend to a smaller radial separation because of the occulter location we used in our observations.}
    \label{fig:classical_contrast_curves}
\end{figure*}

\section{Discussion}\label{discussion}

\subsection{Placing new limits in context}
\label{sec:lim_context}
We are able to better constrain the dust properties of the \gls{epseri} system based on the new surface brightness limits established in this work. Using the fractional luminosity value for the inner disk from \cite{backman_epsilon_2009} (Table 2.), we place an upper limit on the dust albedo of $\omega$ $<$ 0.487, using Equation 5 from \cite{2018ApJ...854...53C}, 
\begin{equation}
\omega=\frac{f_s}{f_e+f_s}    
\end{equation}
where $f_s=3.136\times10^{-5}$ is $ \text{Flux}_{\text{dust}}/\text{Flux}_{\text{star}}$ in the \gls{STIS} bandpass and $f_e=3.3\times10^{-5}$.
To further place our surface brightness results in context, we make use of models generated by the radiative transfer software \texttt{MCFOST} \citep{pinte_2006_montecarlo}, including extensive work done by \cite{wolff2023}, \cite{ertel2020}, and \cite{backman_epsilon_2009}.

We first used the IRS best-fit parameters, that better fit the inner disk, from \cite{wolff2023}, to generate an \texttt{MCFOST} model resembling the structure proposed by \cite{backman_epsilon_2009}. This model has an upper limit of $10^{-7}$(M$_\earth$) in dust mass within 3 AU of \gls{epseri} and a minimum grain size, a$_{\text{min}}$ = 0.5 $\mu$m. We find that this model predicts a surface brightness of 0.012 mJy/arcsec$^2$ at 1\arcsec, which is well below our sensitivity of 0.5 mJy/arcsec$^2$ at 1\arcsec. To further place this model in the context of STIS sensitivity, we used an expression from \cite{debes_pushing_2019} to predict contrast across significantly longer integration times. We found that an integration time of 26.67 hours (i.e. extrapolating to the contrast that could be achieved with a \gls{STIS} program if we use integration times similar to the \gls{Roman}) would yield an improvement of $\sim$4.6 times what we achieved in our program. Accounting for this, we can predict a theoretical surface brightness of 0.11 mJy/arcsec$^2$ achievable by STIS at 1\arcsec, which is still an order of magnitude larger than the brightness predicted by the model informed using parameters from \cite{wolff2023}.

We also consider an alternate model informed by the HOSTS survey, carried out using the Large Binocular Telescope Interferometer (LBTI). They (\citep{ertel2020}), provide a disk flux estimate of 296$\pm$55.6 zodis in the conservative aperture of 27 pixels radius with a scale of 17.9 mas/pixel. Following the procedure in \citep{ertel2020}, we calculated the flux for \gls{epseri} in the N band to be 9325 mJy. Using \texttt{MCFOST}, we  then generated a smooth disk model extending to 1\arcsec with a single power law ($\alpha=-0.45$) using Mie grains of dust mass 1.145$\times10^{-9}$(M$_\earth$),  to agree with the estimated flux from the HOSTS detection. In the HST band-pass, this yields a disk brightness $<$ 0.1 mJy/arcsec$^2$ outside of 0.5\arcsec, which is below our detection limit.

\subsection{Comparison of post-processing approaches}

\gls{NMF} iteratively decomposes the observed \gls{PSF} into a positive, non-orthogonal set of instrumental \gls{PSF} basis modes which avoids the over-subtraction found in the \gls{PCA} family of modeling approaches such as \gls{KLIP}. In contrast to \gls{KLIP}, \gls{NMF} components are sparse and smaller in magnitude, which leads to less over-fitting in the modeling stage, and thus less over-subtraction.
\gls{NMF} also maximizes throughput for extended sources at the expense of some point-source sensitivity and allows us to directly establish a contrast floor and avoid the uncertainty of forward-modeling hypothetical disk geometries to establish a detection limit. In this work, we find that \gls{NMF} has the best disk-focused contrast performance, giving us a contrast of 1.48$\times 10^{-8}$ (corresponding to a surface brightness of 0.59 mJy/arcsec$^2$) at 1\arcsec. 

On the other hand, using \gls{KLIP} provides superior absolute contrast for point sources but suffers from self-subtraction effects, particularly for face-on extended sources. Thus, we performed two \gls{KLIP} subtractions; one focused on an extended source, and one focused on a point source. 
We find that the disk-focused \gls{KLIP} reduction gives us a contrast of 2.25$\times 10^{-8}$ (corresponding to a surface brightness of 0.89 mJy/arcsec$^2$) at 1\arcsec. The planet-focused \gls{KLIP} reduction gives us the best overall contrast of 1.26$\times 10^{-8}$ (corresponding to a surface brightness of 0.5 mJy/arcsec$^2$) at 1\arcsec. The disk-optimized \gls{KLIP} contrast curve (Fig .\ref{fig:klip_contrast_curves}) has slightly worse contrast performance than presented since we calibrated our contrast curve by injecting planetary companions. With performance already worse than \gls{NMF}, we elected to not re-calibrate by injecting a disk model, because that would result in worse performance regardless of the model used. 

We also present a classical RDI subtraction approach, i.e. subtracting the PSF of the reference from the PSF of the target, as a baseline for comparison against NMF and KLIP. We find that classical \gls{RDI} has the worst contrast performance of the three post-processing approaches used in this paper. We achieve a contrast of 4.47$\times 10^{-8}$ (corresponding to a surface brightness of 0.89 mJy/arcsec$^2$) at 1\arcsec.

\subsection{NMF using a broader PSF library}

Building upon the analysis presented in \cite{skpmgpunmf2023}, we also investigated the contrast limits that could be achieved using the entire \gls{STIS} library of frames as our \gls{RDI} \gls{PSF} library, totaling 4796 frames. To produce this curve, we used 90 components constructed from the 10\% closest reference frames, calculated using the Euclidean norm. Excluding \gls{deleri} from the library leads to an order of magnitude worse contrast performance compared to our \gls{NMF} curve produced using only \gls{deleri}. We note, however, that at lower angular separations, it closely approaches the classical \gls{RDI} curve. We posit that in the absence of a good reference source, this is a viable approach to performing \gls{RDI} on a disk-specific dataset. We also postulate that this contrast performance could be improved with better masking of the diffraction spikes produced by the \gls{STIS} coronagraph. Currently, masking is done manually on a frame-by-frame basis, and the number of individual frames using the WEDGE A 1.0 occulter position (4796 frames) makes this an untenable approach. Previous approaches to using an entire instrument archive to build a \gls{PSF} library to perform \gls{RDI} used a simpler mask \citep{Xie_2022} with better results. We also constructed another library with the entire library of frames (including \gls{deleri}) and found that the contrast performance was identical to using \gls{deleri} alone. This was expected since \gls{deleri} constituted most of the closest frames selected by the Euclidean norm measure. We can compare our results to a similar RDI analysis of the SPHERE archive performed by \cite{Xie_2022}, but are unable to reproduce their finding where they show that more references in the library lead to better subtractions overall. This is not surprising, however, given the differences in instruments; the \gls{STIS} CCD is very sensitive to variations in stellar chromaticity. We therefore elected to use only \gls{deleri} as our source to perform \gls{RDI} to save on computational resources. 

\subsection{Current and future observations of \gls{epseri}}

Although we were unable to resolve the inner disk, current and upcoming instruments will likely be able to resolve the disk components of \gls{epseri} in both scattered-light and mid-IR thermal emission. 

The \gls{epseri} system is planned to be observed by the \gls{JWST}, using both the \gls{jwstMIRI} and the \gls{jwstNIRCam} instruments (GTO 1193). The \gls{jwstMIRI} observations are primarily aimed at thermally imaging the outer rings and structures affected by scattered light while the \gls{jwstNIRCam} observations aim to search for massive planets at both a small \gls{IWA} (0.64\arcsec) and large \gls{OWA} (up to 2.2 \arcmin). These observations will likely resolve warm asteroid belt analogs around \gls{epseri} and further our understanding of small grains that contribute to the halo around the system. These observations can potentially provide the first resolved constraints on the disk morphology. 

The Nancy Grace \gls{Roman} Coronagraph Instrument, expected to be launched in 2027, will facilitate scattered-light observations of fainter debris disks (reaching a contrast $\sim10^{-8}$) and probe the dust closer ($\sim$ 1 AU) to the nearby stars \citep{poberezhskiy2022roman,poberezhskiy2021roman,kasdin2020nancy}. The \gls{Roman} Coronagraph supports narrow field observations with the Hybrid Lyot Coronagraph (HLC) at 575 nm and wide field observations with the Shared Pupil Coronagraph (SPC) at 825 nm.  As the \gls{Roman} Coronagraph will be able to probe smaller angular separations with higher contrast than previously achievable, it is expected to detect the scattered light from the unresolved inner disk of \gls{epseri}.  
To simulate observations of the inner \gls{epseri} debris ring through the \gls{Roman} Coronagraph, we first create disk models using the radiative transfer software \texttt{MCFOST} \citep{pinte_2006_montecarlo}. We generated disk models informed by the properties determined in \cite{su_inner_2017,backman_epsilon_2009, wolff2023} and LBTI HOSTS measurements \citep{ertel2020}. The models using the Mie (compact spherical)  grains estimate a disk brightness much greater $>$ 0.5 mJy/arcsec$^2$ outside of 1\arcsec for \cite{backman_epsilon_2009} and \cite{wolff2023} and could have been easily detected in our HST observations. Thus, we created two ring models with properties from \cite{backman_epsilon_2009} and  \cite{wolff2023} using distributed hollow spheres (DHS) grains where the disk brightness is $\sim$ 0.3 mJy/arcsec$^2$ (disk properties provided in Table \ref{table:disk_prop}) and one smooth disk model, informed by the LBTI HOSTS measurements from \cite{ertel2020}.
Since the observed brightness of an IR-excess inferred disk is highly degenerate with grain properties and geometry, we scaled the disk brightness by 0.168mJy/arcsec$^2$ for all the models to the contrast level of 2$\times 10^{-8}$ derived from NMF reduction contrast curve shown in Figure \ref{fig:classical_contrast_curves}.

\begin{figure*}[h]
    \centering
    \includegraphics[width=1\textwidth]{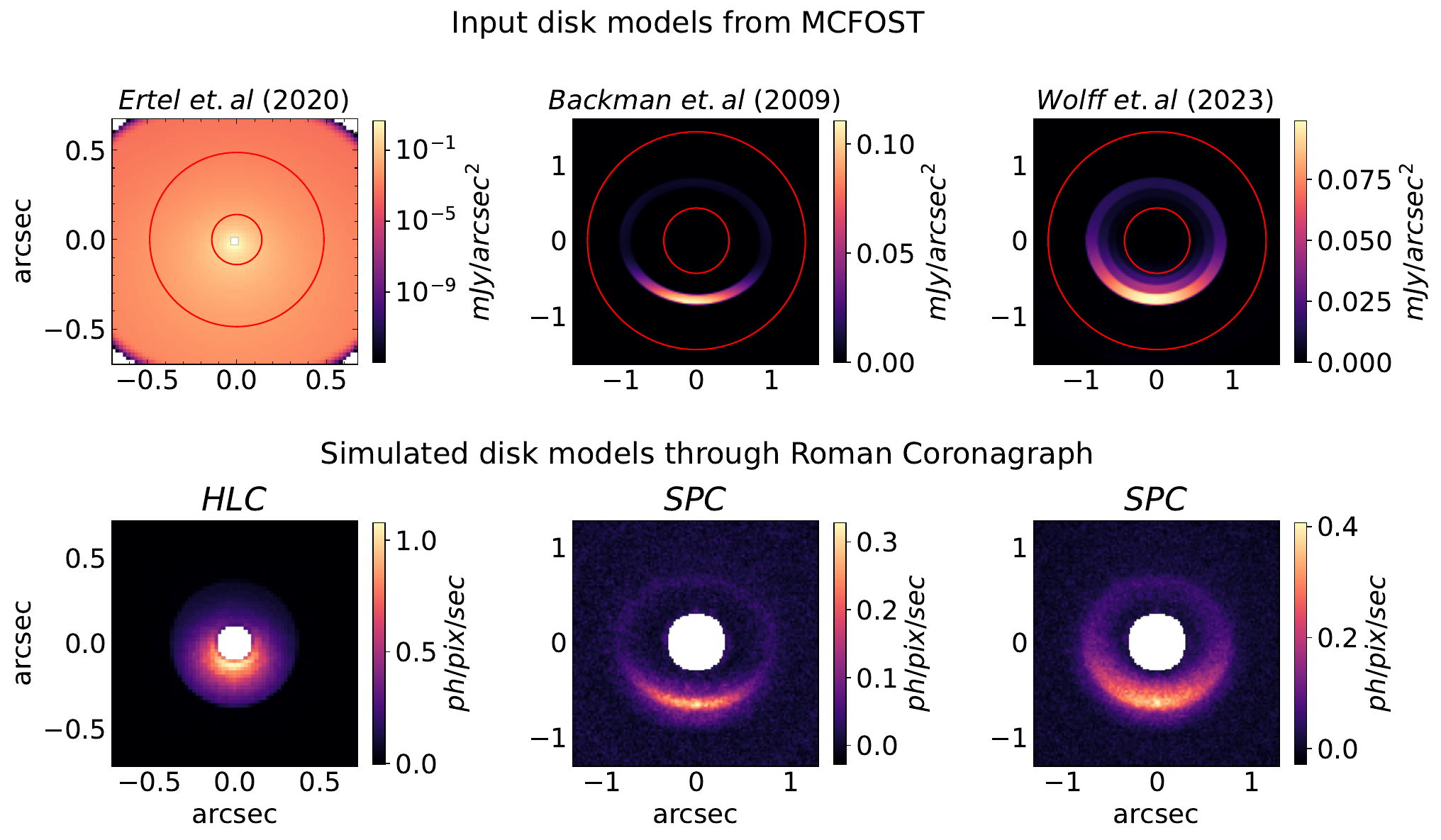}
    \caption{The simulated disk models of \gls{epseri} through the \gls{Roman} Coronagraph Instrument. We use the LBTI HOSTS measurements from \cite{ertel2020} for the smooth disk model and disk properties from \cite{backman_epsilon_2009} and \cite{wolff2023} for the two ring models. The disks are processed through the HLC and SPC mode of the \gls{Roman} Coronagraph instrument using realistic noise and speckle background levels from OS9 (for HLC) and OS11 (for SPC) for a 26.67 hours exposure. The IWA and OWA for the coronagraphs are marked in red for each of the input disk models. The \gls{Roman} Coronagraph will achieve deeper contrast at smaller separations with both modes than we have achieved with \gls{STIS}. }
    \label{eps-eri-roman-cgi}
\end{figure*}

We simulate the \cite{ertel2020} model through the HLC mode and \cite{backman_epsilon_2009} and \cite{wolff2023} models through the SPC mode of the \gls{Roman} Coronagraph as the ring models fall outside the OWA of the HLC mode. We incorporated various noise factors from jitter, speckles, and the EMCCD \citep{nemati2020photon} using the  \href{https://roman.ipac.caltech.edu/sims/Coronagraph_public_images.html#CGI_OS9}{``OS9''} and  \href{https://roman.ipac.caltech.edu/sims/Coronagraph_public_images.html\#Coronagraph_OS11_SPC_Modes}{``OS11''} simulations for HLC and SPC modes, respectively. The various steps involved in the simulation process are explained in more detail in  \cite{anche2023simulation}. The input disk models and the corresponding simulated output disk models are shown in Figure \ref{eps-eri-roman-cgi}. As shown in the lower panel of Figure \ref{eps-eri-roman-cgi}, the inner dust ring around \gls{epseri} is detectable in scattered light with both modes of the \gls{Roman} Coronagraph, depending on the true  size/density distribution of the disk.  
We note the exposure time of these \gls{Roman} simulations is 26.67 hours vs the approximately 0.5 hour integration time here (see \ref{sec:lim_context}); however, the optimized \gls{Roman} coronagraphs are expected to have a much lower photon noise limit and to be insensitive to stellar color mismatch \citep{ygouf_roman_2021}.

\begin{table*}[h]
\begin{center}
\begin{tabular}{llll}
\hline
& Ertel (2020) & Backman (2009) & Wolff (2023)\\ \hline
Dust morphology & Continous & Ring & Ring\\
Disk extent (AU)& 0.1-3     & 3    & 0-3 \\
Dust Mass (M$_\earth$)& 1.149$\times10^{-9}$ & 1.80$\times10^{-7}$  & 9.81$\times10^{-8}$ \\
Minimum grain size-$a_{min}$($\mu$m) & 0.1                  & 3                    & 1                                 \\
Maximum grain size-$a_{max}$($\mu$m) & 1000                 & 3                    & 10000                             \\
Power-law of grain size distribution & 3.65                 & 3.5                  & 3.5                               \\
Surface density exponent  & -0.45 ($\alpha)$                 & 5 ($\alpha_{in}$), -5 ($\alpha_{out}$)                 &   5 ($\alpha_{in}$), -5 ($\alpha_{out}$)                             \\
Grain composition                    & 100\% astrosilicates & 100\% astrosilicates & 60\% astrosilicates               \\
                                     &                      &                      & 40\% organic  refractory material \\
                              \hline  

\end{tabular}
\caption{Summary of disk properties used in modeling the \textit{Ertel} \citep{ertel2020}, \textit{Backman} \citep{backman_epsilon_2009}, and \textit{Wolff} \citep{wolff2023} models using \texttt{MCFOST}. The models generated using these properties were then propagated through the \gls{Roman} Coronagraph. The grain composition for the \textit{Wolff} model was obtained from \cite{ballering2016comprehensive}}.
\label{table:disk_prop}
\end{center}
\end{table*}

\section{Summary}
We have presented new limits on the surface brightness in scattered light of an inner debris disk around the \gls{epseri} system using the \gls{STIS} coronagraph on \gls{HST}. To achieve a small \gls{IWA}, we used the WEDGE A 1.0 occulter location on the coronagraph and observed an effective \gls{IWA} of 0.58\arcsec. Although most of our observed \gls{epseri} frames had the telescope unable to successfully achieve \gls{FGS} lock, we account for these "bad frames" by calculating the standard deviation of all observed frames and discarding the worst 10\%. We applied three post-processing approaches; \gls{NMF}, \gls{KLIP}, and classical \gls{RDI} to attempt imaging the disk and achieve the highest possible contrast. Of these approaches, we find that \gls{NMF} gave us the best contrast performance for an extended source. We observed that an extension of our \gls{NMF} reduction approach by using the entire library of \gls{STIS} coronagraphic frames had negligible impact on final contrast while using a library without any direct reference observations also yielded usable results. We also simulated three models of the \gls{epseri} inner disk using parameters from \cite{backman_epsilon_2009}, \cite{wolff2023}, and \cite{ertel2020}, and found that our measurements were over the sensitivity limit predicted by these models. Additionally, using \gls{ir} measurements from \cite{backman_epsilon_2009}, we have placed an upper limit on the dust albedo, $\omega$ $<$ 0.487. JWST observations are likely to resolve the spatial extent of the disk which will better constrain disk parameters for future scattered light observations with observatories such as \gls{Roman}, which we expect to have a significantly deeper sensitivity limit.

\section*{Acknowledgements}

We thank Schuyler Wolff and Andr\'as G\'asp\'ar for valuable discussions on contrast calculations and consultations on using the AUTOFILET program. We also extend special thanks to Karl Misselt for sharing his expertise in the Fortran programming language in debugging the AUTOFILET program. Additionally, we would like to extend our thanks to Steve Ertel for help with interpreting the HOSTS survey measurements to include in our modeling efforts. We would also like to thank the anonymous referee for their assistance in making this a more comprehensive and well rounded piece of work.

Based on observations made with the NASA/ESA \textit{Hubble Space Telescope}, obtained from the data archive at the Space Telescope Science Institute, which is operated by the Association of Universities for Research in Astronomy, Inc., under NASA contract NAS 5-26555. These observations are associated with program GO 15217. The specific observations analyzed can be accessed can be accessed via \dataset[DOI: 10.17909/gxm6-5x86]{https://doi.org/10.17909/gxm6-5x86}. STScI is operated by the Association of Universities for Research in Astronomy, Inc. under NASA contract NAS 5-26555. 
Support for program HST-GO-15217 was provided by NASA through a grant from the Space Telescope Science Institute, which is operated by the Association of Universities for Research in Astronomy, Inc., under NASA contract NAS 5-26555.

Portions of this research were supported by funding from the Technology Research Initiative Fund (TRIF) of the Arizona Board of Regents and by generous anonymous philanthropic donations to the Steward Observatory of the College of Science at the University of Arizona.

This material is based upon High Performance Computing (HPC) resources supported by the University of Arizona TRIF, UITS, and Research, Innovation, and Impact (RII) and maintained by the UArizona Research Technologies department.

\facilities{\textit{HST} (STIS), Barbara A. Mikulski Archive for Space Telescopes} 
\software{This research made use of community-developed core Python packages, including: 
Astroquery \citep{adam_ginsburg_astropy/astroquery_2018}, 
Astropy \citep{the_astropy_collaboration_astropy_2013, 2018AJ....156..123A, 2022ApJ...935..167A}, 
ccdproc \citep{ccdproc},
CuPy \citep{cupy_learningsys2017},
Matplotlib \citep{hunter_matplotlib_2007}, 
NumPy \citep{harris2020array},
Pandas \citep{reback2020pandas, mckinney2010data},
Poppy \citep{poppy},
PyKLIP \citep{wang_2015_pyklip},
centerRadon \citep{centerRadon}
SciPy \citep{jones_scipy_2001}, 
Multiprocess \citep{multiprocess_main, multiprocess_pathos}, and
the IPython Interactive Computing architecture \citep{perez_ipython_2007}.}
Additional data analyses were done using IDL (Exelis Visual Information Solutions, Boulder, Colorado). This research has made use of the SIMBAD database,
operated at CDS, Strasbourg, France \citep{Wenger_2000}.


\appendix

\section{Constraints on substellar companions around $\epsilon$ Eri}
\label{appendix:planet}

\begin{figure*}[ht]
    \centering
    \includegraphics[width=0.95\textwidth]{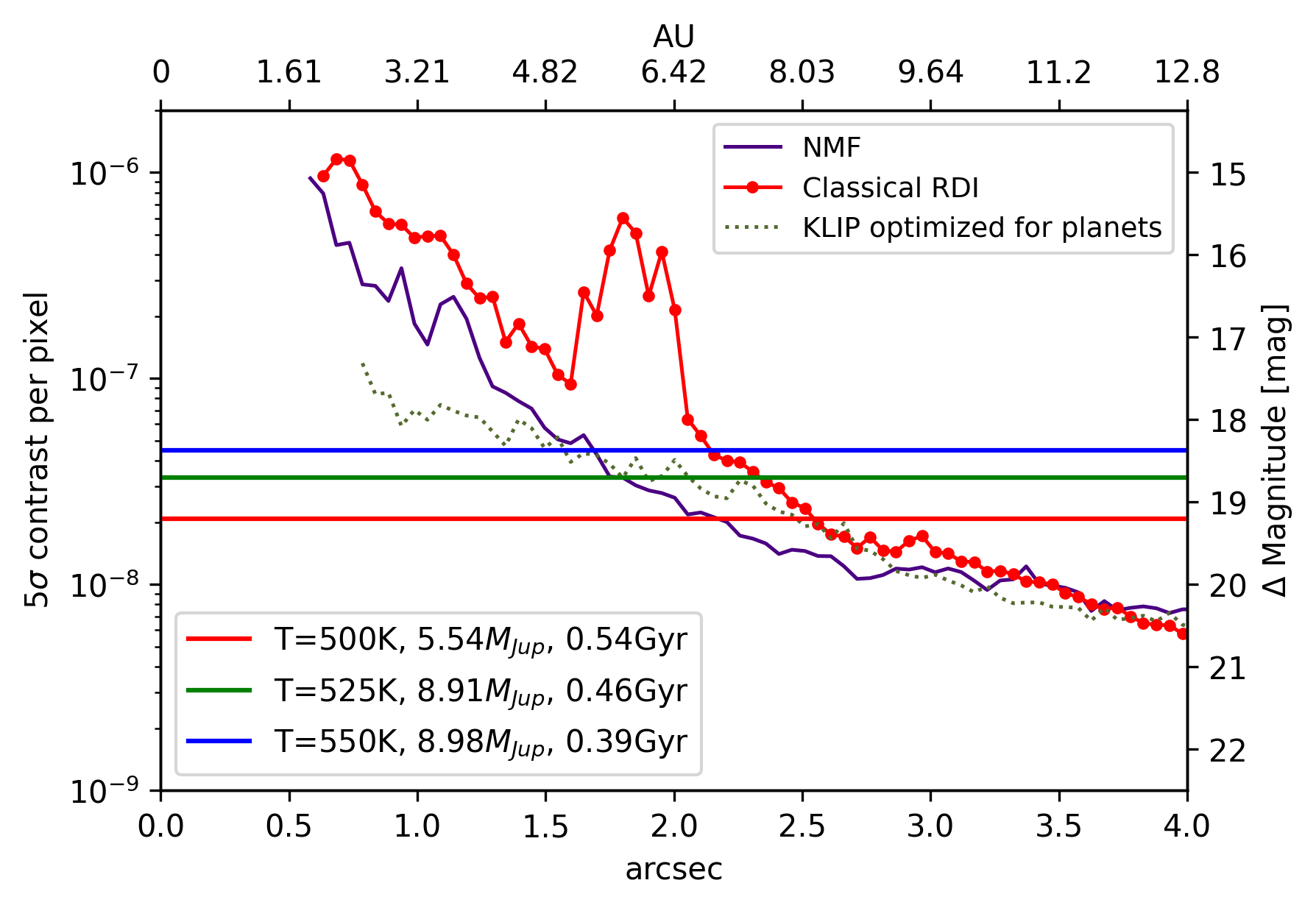}
    \caption{Contrast curves from three different post-processing approaches plotted against predicted planetary masses (using the Sonora-Bobcat planetary evolution models \citep{sonora_bobcat1}) in units of M$_\text{Jup}$. We use $5\sigma$ contrast instead of MAD contrast for this plot to maintain consistency with published exoplanet literature. This plot assumes purely emissive light contrast and does not take reflected light from the planet into effect.}
    \label{fig:nmf_mass}
\end{figure*}

We used the Sonora Bobcat models \citep{sonora_bobcat1} considering three different temperatures (500K, 525K, and 550K), and considering solar metallicity with model spectra that fall within the expected age of the \gls{epseri} system. We then used the model spectra to determine count rates for the STIS CCD using the Hubble ETC and then calculated contrast. At the highest temperature we considered (550K), we calculated an upper detection limit on the mass of substellar companions of 8.98 M$_{\text{Jup}}$. This corresponds to a contrast of 4.49$\times 10^{-8}$, which is slightly less than 6.30$\times 10^{-8}$, the contrast we achieve at 1\arcsec.
 We derived this using the planet-focused \gls{KLIP} reduction, which provides the best absolute contrast at small angular separations. We note that these detection limits do not take into account any reflected light, since that would require assumptions on radius and atmospheric composition that are not known. Since the \cite{mawet_deep_2018} observations constrain companion mass to $\sim$1 M$_{\text{Jup}}$ at $\sim$1\arcsec, a non-detection is not surprising in this case, and even accounting for reflected light would not overcome the order of magnitude difference between prior constraints and our detection limits. Detection limits for three different post-processing approaches are plotted in Figure \ref{fig:nmf_mass}. A consequence of using standard deviation to calculate contrast is reflected in the Classical \gls{RDI} plot. Since our data has a larger occurrence of outliers, the sensitivity of standard deviation leads to worse contrast calculations between 1.5\arcsec to 2\arcsec. 

\section{Upper limits on disk flux}
\label{sec:appendB}
\begin{figure*}[h!]
    \centering
    \includegraphics[width=0.95\textwidth]{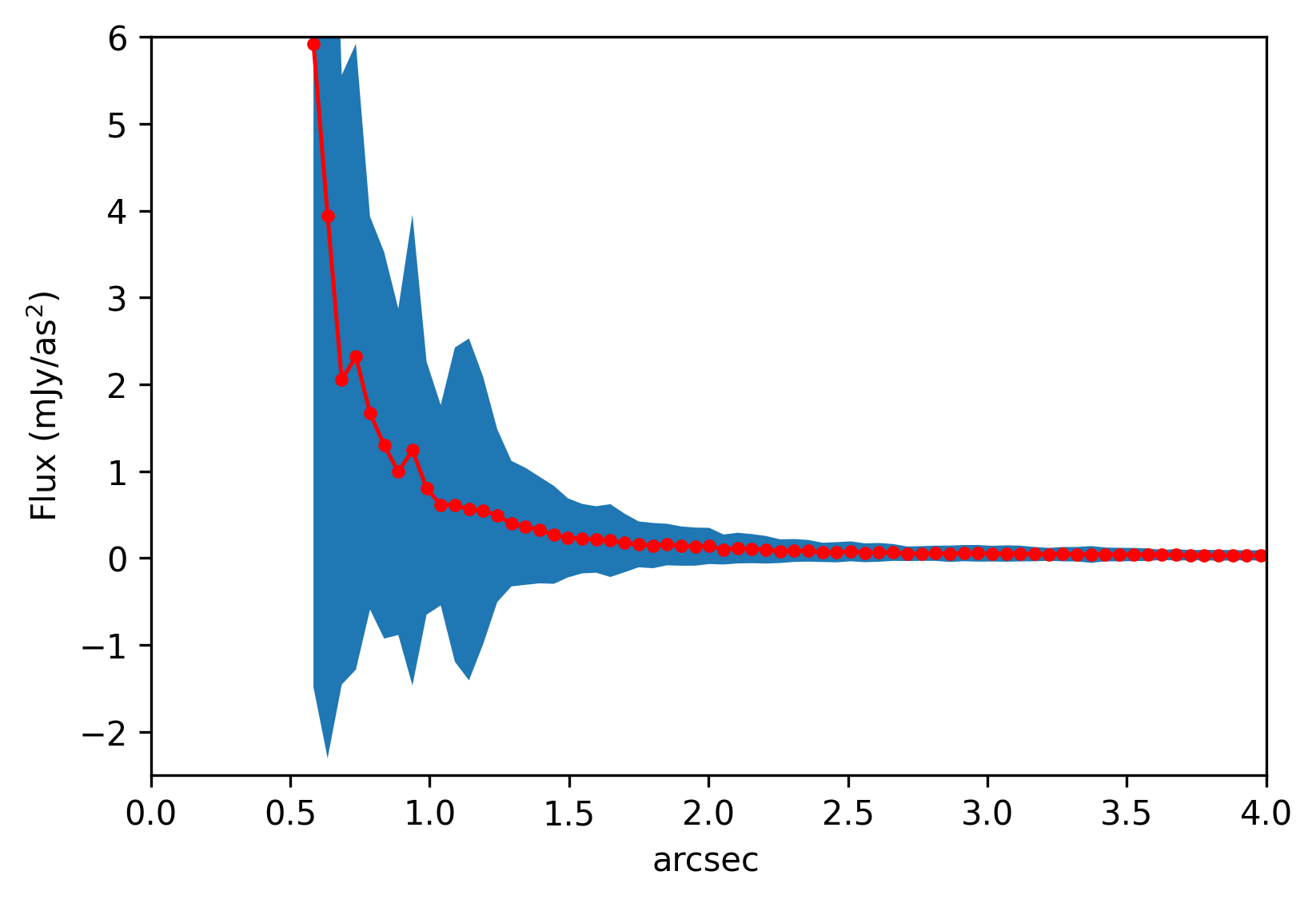}
    \caption{Flux profile of the \gls{epseri} system using \gls{NMF} in red. The shaded blue region represents the 1$\sigma$ uncertainty in the measurement. Consistent with our derived SNR map, our result is indicative of a non-detection.}
    \label{fig:flux_profile}
\end{figure*}

To quantify the absolute lowest sensitivity we achieved from our reductions, we measure an upper limit on the radial flux profile of our NMF reduction, which achieved the highest contrast in the region where the inner disk would be, in Figure \ref{fig:flux_profile}. We present the 1$\sigma$ uncertainty on the measurement as a shaded region, determined from the noise map used in the creation of the right panel of Figure \ref{fig:nmf_final}. Consistent with the SNR map of our NMF reduction, our measurement is consistent with a non-detection.

\bibliography{epsEri}
\end{CJK*}
\end{document}